\documentstyle[12pt,aasms4]{article}
\slugcomment{\shortstack[r]{
To be submitted to {\sl The Astronomical Journal}
}}

\begin{document}

\title
{The Faint End of the Galaxy Luminosity Function In Abell 426 and 539}

\author
{Roberto De Propris\altaffilmark{1}\altaffilmark{2}}

\affil
{Jet Propulsion Laboratory, California Institute of Technology, Pasadena,
California, 91109; e-mail: propris@coma.jpl.nasa.gov}

\altaffiltext{1}
{NRC Resident Research Associate}

\author
{Christopher J. Pritchet\altaffilmark{2}}

\altaffiltext{2}
{Guest Astronomer, Canada-France-Hawaii Telescope, which is operated
in conjunction by the National Research Council of Canada, Le Centre
National de la Recherche Scientifique of France and the University of
Hawaii}

\affil
{Department of Physics and Astronomy, University of Victoria, PO Box 3055,
Victoria, BC, V8W 3P6, Canada; e-mail: pritchet@clam.phys.uvic.ca}

\begin{abstract}

We derive $I$ band luminosity functions for galaxies in Abell 426 (Perseus
and Abell 539, two rich, low galactic latitude clusters at moderate
redshift. Cluster members are selected via the color-magnitude relation
for bright galaxies. We find $\alpha=-1.56 \pm 0.07$ for Perseus over the
range $-19.4 < M_I < -13.4$ ($15 < I < 21$) and $\alpha=-1.42 \pm 0.14$ 
for $-18.5 < M_I < -14$ ($17 < I < 21.5$) for A539. These LF's are similar
to those derived in Virgo and Fornax, weakly supporting claims for the
existence of a universal luminosity function for galaxies in clusters.
\end{abstract}


\section
{Introduction}
Dwarfs are the most common type of galaxy in the Universe; yet, because
of their low luminosity and surface brightness our understanding of these
objects is rather limited. Only recently has it become possible to obtain
photometry for statistically significant samples of dwarfs, thanks to the
availability of wide field CCD detectors on large telescopes.\par
Luminosity functions (LF) for galaxies are a powerful tool to examine
galaxy bulk properties, especially where little detailed information is
available, as in the case of dwarf galaxies. LF's can be obtained with
a modest amount of observing time, are relatively easily corrected
for incompleteness, and possess considerable physical meaning. For
example, the litmus test of any theory of galaxy formation is how well
it reproduces the observed LF.\par
Clusters of galaxies at moderate redshift ($z < 0.1$) are one of the best
environments to study the dwarf LF, because of their high density and
the possibility of dealing with contamination statistically.  The fact
that the nearest clusters are about 5 times as distant as Virgo or Fornax
is actually an advantage for dwarf galaxy completeness.  This is because
low surface brightness galaxies tend to become easier to detect as their
angular size decreases. In addition, the ratio of target objects to
background galaxies is roughly independent of the distance modulus of the
cluster under consideration, but since the angular correlations of fainter
galaxies are smaller, field-to-field fluctuations in the background counts
are decreased. This gives us greater confidence in the derived LF slope.
In addition, it is possible to survey a number of environments by choosing
clusters of different densities, richness, Bautz-Morgan class, Rood-Sastry
type or evolutionary stage. These considerations were the basis of the
pioneering survey of the Virgo cluster by Sandage et al.\ (1985).\par
Recent investigations have revealed large numbers of dwarfs to be present
in clusters, whose LF slopes are sometimes as steep as $\alpha =-2$ (where
$\alpha$ is the slope $dN/dL$: $\alpha=-2$ corresponds to $d \log N/dm=+0.4$,
whereas $\alpha=-1$ is flat in a plot of $\log N$ vs. magnitude: N is the
number of galaxies).\\
We tabulate a compilation of recent results in Table 1: column 1 identifies
the cluster, column 2 presents the $\alpha$ values derived and their range,
column 3 gives the magnitude limits of the fit and column 4 identifies the
reference. A comparison of these results is somewhat difficult, since
the assumption that the LF follows a Schechter (1976) function is no
longer tenable. Trentham (1998a) shows that LF's tend to drop rapidly
at bright ($M_B < -21$) luminosities, flatten for $-18 < M_B < -16$ and
then rise for fainter galaxies, with slopes $-1.3 < \alpha < -1.8$.
The above `shape' of the LF may be at the origin of the `dips' and `bumps'
seen in the Coma cluster (Lobo et al.\ 1997) and in A2029 (De Propris et 
al.\ 1998b). One further complication is that removal of background 
galaxies from the cluster counts is carried out in a number of different
ways: Bernstein et al.\ (1995) and Trentham (1997, 1998a,b) compute $R$
band counts and their variance from observations of background fields.
De Propris et al.\ (1995) use the correlation function to estimate
the field-to-field variance in their $B$ and $R$ counts, but they normalize
these counts to observations of a background field. Secker (1996) uses
the $B-R$ color magnitude relation to isolate cluster members, whereas
Lobo et al.\ (1997) adopt $V$ band counts from the ESO-Sculptor Faint 
Galaxy Redshift Survey (Arnouts et al.\ 1997). The slopes reported by these
authors vary from `shallow' ($\alpha > -1.5$) to very steep ($\alpha \sim
-2$), although over different magnitude ranges. Determination of accurate
background counts is crucial for estimates of the LF, since the slope of
the background counts, in our formalism, is equivalent to $\alpha \sim -2$
and would therefore tend to artificially steepen derived LF's.\\
The case of Coma is particularly interesting: Lobo et al.\ (1997)
claim a very steep $V$ band LF ($\alpha \sim -1.8$) whose slope appears
to increase outward from the cluster core (Sekiguchi et al.\ 1998). De
Propris et al.\ (1998a) show that this steep LF persists in the infrared
$H$ band, and is therefore a feature of the mass function of galaxies,
in accordance with the hypothesis that stabursts drive the optical LF
(Hogg \& Phinney 1997), and a steep LF is consistent with a flat mass
function for dwarf galaxies.\\
Lobo et al.\ (1997) and Sekiguchi et al.\ (1998) sample their LF to
$M_V < -14.75$ and $M_B < -16$ respectively. 
De Propris et al.\ (1998a) find a steep upturn for $M_H < -19$, approximately
corresponding to $M_B < -16$, whereas Trentham (1998b) reports a shallower
slope over this magnitude range. Further, the `depletion' observed by Lobo et
al.\ (1997) near NGC4889 and NGC4874 is limited to relatively bright galaxies,
whereas Trentham (1998b) reports an increase at fainter magnitudes. This is
consistent with earlier observations of A262 where dwarfs appeared to be
concentrated in the cluster core (Trentham 1997). It is indeed possible
for the dwarf-rich  luminosity function to be ubiquitous, with examples 
found even in the field (Marzke et al.\ 1994; Loveday 1997). On the other 
hand, a number of authors disagree and report optical LF slopes $\alpha
\simeq -1.5$ (Bernstein et al.\ 1995; Secker 1996; Lopez-Cruz et al.
1997).\par
It has often been suggested that the cluster environment should strongly
affect the fragile dwarf galaxies. Lobo et al.\ (1997) find that the
LF in the immediate vicinity of NGC4874 (the cD galaxy in Coma) is 
shallower. Lopez-Cruz et al.\ (1997) have proposed that cD galaxies
grow at the expense of their dwarf satellites. Trentham (1997) also
suggests that the slope of the LF decreases as clusters evolve and 
dwarfs are subsumed by giants. It is nevertheless possible for the
cluster environment to favor the formation or preservation of dwarf
galaxies. Dwarfs may be formed during mergers (Krivitsky \& Kontorovich
1997) and are seen to form in tidal tails (Mirabel et al.\ 1992; Yochida
et al.\ 1994), with steep LF slopes (Hunsberger et al.\ 1996). Silk et
al.\ (1987) have suggested that dwarfs may accrete gas from the intracluster
medium and fuel further bursts of star formation.\par
Here, we wish to entertain the hypothesis that cluster evolution is
responsible for these effects. We have therefore selected targets
using the X-ray classification scheme by Jones \& Forman (1984).
In this model, clusters evolve from loose, spiral-rich, X-ray faint
and low density objects, to dense, X-ray luminous and elliptical rich
systems. An additional benefit of this scheme is that it allows us to
consider the role of gas density (Babul \& Rees 1992).\par
In this paper, we detail observations of two clusters: Abell 426 (Perseus)
and Abell 539. Perseus is an old cluster, with high X-ray luminosity
and a small core radius. Galaxies are distributed on a large filament,
which appears to be aligned with surrounding structure. Many of its bright
members show signs of nuclear activity (e.g., NGC1275 is a Seyfert 2 AGN,
IC310 is a WAT radio source).  Abell 539 is a rich cluster, dominated by
a chain of elliptical galaxies and believed to be undergoing a collision
(Ulmer et al.  1992). In the Jones \& Forman (1984) classification,
this object is slightly more evolved than Virgo. Observing these objects
we hope to elucidate the role played by cluster evolution on the dwarf
galaxy LF, and to enlarge the sample of well studied clusters.\par
In this paper, Section 2 presents our observations and data reduction.
Section 3 introduces our analysis. Section 4 details our results, which
we discuss in section 5. Our conclusions are summarized in section 6. We
choose a cosmology with $H_0=70$ and $q_0=0.1$ throughout

\section
{Observations and Data Reduction}
Observations were taken during the nights of December 04-06, 1994 at
the prime focus of the Canada-France-Hawaii 3.6m Telescope, using a $4K
\times 4K$ mosaic CCD camera (Cuillandre et al.\ 1996). Two of the nights
were photometric, although the seeing was generally poor (about 
$1''$) for CFHT, whereas the third had to be abandoned because of very poor
weather. In the version used in this paper, and now superseded, MOCAM
was equipped with four engineering-grade chips. The cosmetic quality 
of these detectors was generally satisfactory, except for one chip 
which was found to be affected by a large defect. Readout times were,
however, very long (about 10 minutes each) and this limited our
data collecting efficiency considerably. The pixel scale was about 
$0.21''$ and the total field of view $14' \times 14'$.\par
We observed two fields, centered on the giant elliptical galaxies
NGC1275 (Abell 426) and UGC3274 (Abell 539), in $V$ and $I$.
The total exposure times were 1200s in both filters, suitably split
to aid in the removal of cosmic rays.\par
Data reduction was carried out via the usual techniques of trimming,
debiasing and flatfielding. The CCD mosaic data was split into its
four component chips before these steps. The cosmetic defect seen to
exist in one of the chips proved impervious to any processing procedure
and we were forced to remove this chip from all further analysis. This
leaves us with a total observed area of 147 square arcminutes.\par
Because of the long readout times we found it impossible to obtain
twilight sky flatfields. Our images were strongly contaminated by
bright galaxies and by saturated stars (because of the low galactic
latitude of our targets) and could not be used to build sky flats.
Eventually, we decided to use dome flatfields. This may reduce the
accuracy of our photometry, but the additional small uncertainty in
our photometry is not relevant for our studies of the luminosity
function.\par
Finally, we coadded all data for each cluster to improve our signal 
to noise ratio. The final images are analyzed as discussed below.

\section
{Analysis}

It is immediately apparent, from a cursory examination, that our
images are dominated by the two central cluster giants (Figures 1
and 2). To remove diffuse light from these objects and expose
galaxies `hidden' beneath their bulk, we have modelled these galaxies
using the IRAF STSDAS routines {\tt ellipse} and {\tt bmodel}, and
removed them from the images.\par
To detect objects, we convolved each image with a lowered Gaussian,
having Full Width at Half Maximum (FWHM) equivalent to the stellar
FWHM, and with a kernel about four times as large. Objects which are
above a $4\sigma$ level from the mean sky are included in our
catalog for later photometry. Here $\sigma$ refers to the noise in
the sky {\it after} convolution with the lowered Gaussian.\par
For each object, we computed Kron (1980) image moments $r_1$ and $r_{-2}$.
The first parameter, $r_1$, is a `size' estimator and is used for
photometry, which is carried out in $2r_1$ (radius) apertures. Infante
(1987) and Infante \& Pritchet (1992) show that this procedure encloses
most of the light from each object. Naturally, this procedure is not
employed for bright members, where the single aperture which we use to
measure the image moment is not sufficient. For these objects, we use
surface brightness profiles to estimate the best aperture (i.e., one
which includes as much light as the Kron apertures used for fainter
objects). The $r_{-2}$ parameter is used as a measure of `compactness'
in order to discriminate between stars and galaxies. In Figures 3 and
4 we plot $r_{-2}$ vs. $I$: a `stellar' sequence at a constant $r_{-2}$
is clearly visible. This sequence inflects upward at bright magnitudes
because of saturation, but these objects are easily identified by 
inspection. We remove stars from our catalogs, which then consist 
solely of galaxies.\par
Photometry is calibrated by observations of stars in M67 by Montgomery
et al.\ (1993). Because of the low galactic latitude of our targets we
need to correct for foreground extinction. We use the reddening maps by
Burstein \& Heiles (1982), transformed to the bands of interest using
Johnson (1965). Our estimates of foreground extinction agree with 
those stated in NED and by Weedman (1975). They are equal to 0.54 and
0.40 magnitudes in $V$ and $I$ (respectively) for Perseus, and 0.18 and
0.13 magnitudes for A539.\par
We select galaxies in the $I$ band, since the LF will be closer
to the underlying mass function. For each object we compute $V-I$ colors
in $3 \times $ FWHM (radius) apertures. We bin data in 0.5 magnitudes bins
to produce galaxy number counts for objects in our fields. These counts
consist of a contribution from cluster members and one from foreground 
and background galaxies. These have to be removed from consideration.
We adopt two methods: one is to remove background counts statistically,
using estimates from the literature (Lilly et al.\ 1991; Gardner et al.\
1996). The other is to use the $V-I$ color-magnitude relation to assess
membership, since cluster members tend to lie on a well-defined ridge 
line. We assume that all objects within $\pm 0.3$ of the color-magnitude
relation shown in Figures 5 and 6 (for A426 and A539 respectively) are
members (Mazure et al.\ 1988).\par
In Figures 7 and 8, we compare counts obtained with these two techniques.
It can be seen that the agreement is satisfactory, if not perfect. We
have used counts from the literature since the presence of structure in
the vicinity of Perseus (as is the case for most clusters) may affect
determinations of the background level from nearby `blank' fields. 
Therefore, counts from the literature may better reflect the true level
of background and foreground contamination. Note that these counts are
{\it not} used for LF determination: they are shown here to demonstrate
that the color-selected sample does not bias our result against blue
cluster members, which are unlikely to exist in the cluster core.\par 
The shape of the LF is not fit by a single power law, a composite
function being necessary (Trentham 1998a). On the other hand, a
power-law, despite its lack of physical meaning, is still a concise
way of summarizing these data and we choose to adopt it in this
paper, even when, as in Figure 10, a power-law fit does not appear
to represent the data very well ($\chi^2/\nu =2.9$).\par 
We show these fits in Figures 9 and 10. For Abell 426 we derive a
slope of $\alpha=-1.56 \pm 0.08$ over $-19.4 < M_I < -13.4$. For Abell 539
we obtain $\alpha =-1.42 \pm 0.14$ for $-18.5 < M_I < -14.0$.\par

\section
{Discussion}

The slopes we derive for A426 and A539 are broadly consistent with 
those found in Virgo (Sandage et al.\ 1985, but see Phillipps et al.\ 1998) 
and the Fornax cluster (Ferguson \& Sandage 1988). Our derived slopes
are only marginally different from each other. Although the slope measured
for A426 is slightly steeper than for A539, the difference is not
statistically significant. Abell 426 is more evolved than A539, on the
Jones \& Forman (1984) scale, as shown by his preponderantly elliptical
population (as opposed to the more spiral-rich A539). Therefore, our
results are mildly inconsistent with the hypothesis that cluster evolution
results in a flatter LF, but the significance of this is low.\par
On the other hand, dwarfs may be created in clusters. NGC1275 is believed
to lie at the center of a large cooling flow, which usually associated
with star formation. Globular clusters are known to have been
formed recently in this galaxy (Holtzman et al.\ 1992). This may have
boosted the LF slope in Perseus. On the other hand, such objects would
likely be blue; our color-selection criterion would exclude most
of these galaxies and we may therefore underestimate the slope of the
LF in Perseus. The good agreement of the color-selected and background
subtracted counts in Figures 5 and 6 suggests that our LF slope is not 
affected by this problem.\par
Note of course that there is some evidence for an environmental effect
in the Coma cluster. The brighter Lobo et al.\ (1997) sample shows
fewer galaxies in the immediate vicinity of the two giants and an increase
at larger clustercentric distances. Fainter galaxies are instead abundant
throughout the cluster and may be concentrated towards the cluster core
(Trentham 1998b): this is similar to what has been found in A262 (Trentham
1997). We should not be affected by these environmental dependencies in 
our analysis.\par
If the above results are correct, there may exist two classes of clusters:
objects with steep dwarf LF's (such as Coma, Abell 2199) and objects with
flatter LF's (Virgo, Perseus). It is not easy, at present, to identify 
the physical mechanisms behind this, since both groups are rather
heterogeneous, including objects as different as Coma and A262 (Trentham
1997) on one side, and Perseus and Virgo on the other. Unfortunately,
as shown in Table 1, it is possible for measurements of the same cluster
to yield different results, so that there is little solid evidence for
a bimodal distribution of LF slopes.\par
Clearly, the subject of dwarf galaxy LF's is still disputed and has
important theoretical consequences. Steep LF's are, as we pointed out,
predicted by Cold Dark Matter scenarios for galaxy formation (e.g.,
White \& Frenk 1991), but these are believed to exist in the field
rather than in clusters (Dekel \& Silk 1986). If the steep LF were
continued to luminosities comparable to those of Carina or Draco,
the mass enclosed in dwarf galaxies would be considerable, especially
if the trend for higher Mass to Luminosity Ratios for fainter objects
(Pryor 1992) continues in distant clusters. This may reduce the need
for a large non baryonic `halo' component to accomodate cluster dynamics.\par
New panoramic detectors and infrared arrays allow accurate LF's and
mass functions to be determined over large fields. These new technologies
allow a wide optical-infrared survey to be undertaken, to clarify the
shape of the mass and luminosity function in clusters.
\section{Conclusions}
We derive $I$ band luminosity functions for galaxies in Abell 426 and
Abell 539. For these clusters we find a good fit to a power-law with
slope $\alpha=-1.56 \pm 0.08$ for Perseus and $-1.42 \pm 0.14$ for A539.
Since A426 is the more evolved of the two clusters, our findings run
counter to the idea that cluster evolution destroys dwarf galaxies:
on the other hand, star formation may be responsible for boosting the
LF slope of dwarf galaxies in this system. 

\acknowledgments

The research described here was carried out by the Jet Propulsion 
Laboratory, California Institute of Technology, under a contract with NASA.

Part of this work was carried out while R. D. P. was a Graduate Student
at the University of Victoria. R. D. P. would like to thank the University
of a Victoria for a Graduate Fellowship, a R. M. Petrie Fellowship and
the Advanced Systems Institute of British Columbia for a Ph. D. entrance
Fellowship. R. D. P. would also like to thank the National Research Council
(USA) for a Resident Research Associateship at the Jet Propulsion Laboratory.
The work of C. J. P. is supported by the Natural Sciences and Engineering
Research Council of Canada (NSERC).\par

We would like to thank Drs. Y. Mellier and J.-C. Cuillandre for their
assistance in the use of MOCAM, and Drs. F. D. A. Hartwick and R. O. Marzke
for helpful discussions. We would also like to thank an anonymous referee
for many useful suggestions, which have substantially improved the scope
of this paper.
\clearpage
\centerline{LIST OF FIGURES}

Figure 1: $I$ band image of the central region of Abell 426 (Perseus)

Figure 2: $I$ band image of the central region of Abell 539

Figure 3: Concentration parameter ($r_{-2}$) vs. $I$ magnitude
for Abell 426 (Figure 1). Objects are plotted as open circles.
The adopted star-galaxy separation value is shown as a thick 
dashed line.

Figure 4: Concentration parameter ($r_{-2}$) vs. $I$ magnitude
for Abell 539 (Figure 2). Objects are plotted as open circles.
The adopted star-galaxy separation value is shown as a thick 
dashed line.

Figure 5: Color-Magnitude diagram for galaxies in A426. The ridge
line (thick solid line) shown is derived from the colors of spectroscopic
members of the cluster (Chincarini \& Rood 1971). Objects within $\pm
0.3$ magnitudes of this line are assumed to be members.

Figure 6: Color-Magnitude diagram for galaxies in A539. Spectroscopic
information is taken from Ostriker et al.\ (1988).

Figure 7: Comparison between color-selected (filled circles) and background
subtracted (filled squares) galaxy counts in Abell 426

Figure 8: As Figure 7, but for A539

Figures 9: Luminosity function, derived from color-selected counts,
for A426.

Figure 10: Luminosity function, derived from color-selected counts, for A539
\acknowledgments

\end{document}